\documentclass[11pt]{article}

\usepackage{amsfonts}
\usepackage{amsmath}
\usepackage{amssymb}
\usepackage{amsthm}
\usepackage{mathrsfs}
\usepackage{psfrag}
\usepackage{color}

\usepackage{graphicx}

\theoremstyle{plain}
\newtheorem{theorem}{Theorem}
\newtheorem{proposition}[theorem]{Proposition}

\newtheorem{lemma}[theorem]{Lemma}

\theoremstyle{definition}

\numberwithin{exercise}{section} \numberwithin{equation}{section}
\numberwithin{theorem}{section} \numberwithin{problem}{section}

\numberwithin{figure}{section}

 \DeclareMathOperator{\diag}{diag}

\newcommand{\bs}[1]{{\boldsymbol{#1}}}

\newcommand{\R}{\mathbf{R}}
\newcommand{\N}{\mathbf{N}}

\begin{document}
\title{Rigorous mathematical analysis of the quasispecies model:\\ From Manfred Eigen to the recent developments}

\author{Alexander S. Bratus$^{1,2}$, Artem S. Novozhilov$^{{3},}$\footnote{Corresponding author: artem.novozhilov@ndus.edu}\,\,, Yuri S. Semenov$^{2}$\\[3mm]
\textit{\normalsize $^\textrm{\emph{1}}$Faculty of Computational Mathematics and Cybernetics,}\\[-1mm]
\textit{\normalsize Lomonosov Moscow State University, Moscow 119992, Russia}\\[2mm]
\textit{\normalsize $^\textrm{\emph{2}}$Applied Mathematics--1, Moscow State University of Railway Engineering,}\\[-1mm]\textit{\normalsize Moscow 127994, Russia}\\[2mm]
\textit{\normalsize $^\textrm{\emph{3}}$Department of Mathematics, North Dakota State University, Fargo, ND 58108, USA}}

\date{}

\maketitle

\begin{abstract}
We review the major progress in the rigorous analysis of the classical quasispecies model that usually comes in two related but different forms: the Eigen model and the Crow--Kimura model. The model itself was formulated almost 50 years ago, and in its stationary form represents an easy to formulate eigenvalue problem. Notwithstanding the simplicity of the problem statement, we still lack full understanding of the behavior of the mean population fitness and the quasispecies distribution for an arbitrary fitness landscape. Our main goal in this review is two-fold: First, to highlight a number of impressive mathematical results, including some of the recent ones, which pertain to the mathematical development of the quasispecies theory. Second, to emphasize that, despite these 50 years of vigorous research, there are still very natural both biological and mathematical questions that remain to be addressed within the quasispecies framework. Our hope is that at least some of the approaches we review in this text can be of help for anyone embarking on further analysis of the quasispecies model.


\end{abstract}

\section{Introduction}
In 1971 Manfred Eigen published a groundbreaking paper ``Selforganization of matter and the evolution of biological macromolecules''\cite{eigen1971sma}, in which he considered various aspects of the problem of the origin of life. As a part of his comprehensive approach he introduced a system of ordinary differential equations, nowadays commonly called \textit{the quasispecies model}, which since then became one of the classical models in the field of mathematical biology. Despite of its intrinsic simplicity (the model is ``almost'' linear) and despite almost fifty years of vigourous research, there are still open and deep mathematical questions about this model. Our goal in this review is to discuss both classical and relatively recent progress in the mathematical aspects of the analysis of the quasispecies model and also highlight some open problems. We exclusively concentrate on the mathematical side of the story and refer the interested reader to the earlier reviews \cite{baake1999,eigen1988mqs,jainkrug2007,schuster2015quasispecies} and to the whole recent volume \cite{domingo2015} devoted to the various other sides of the quasispecies theory.

While the original quasispecies model was formulated in continuous time we start with a more natural discrete time settings. We consider a population of individuals such that there are $l$ different types. Let $n_i(t)$ denote the number of individuals of type $i$ at the time moment $t$. The individual reproduction success is described by the constant fitness coefficients $w_i\geq 0$, which we put together into the diagonal matrix $\bs W=\diag(w_1,\ldots, w_l)$ or vector $\bs w=(w_1,\ldots,w_l)^\top\in \R^l$, to both of which we refer as \textit{fitness landscape}. Moreover, the reproduction is error prone such that the probability that an individual of type $j$ begets an individual of type $i$ is given by $q_{ij}\in[0,1]$, and hence we have the stochastic mutation matrix $\bs Q=[q_{ij}]$, where $q_{ii}=1-\sum_{j=1,j\neq i}^l q_{ij}$ is the probability of faithful reproduction. Now simple bookkeeping yields the following recurrence equation
\begin{equation}\label{eq:1:1}
    n_i(t+1)=\sum_{j=1}^l w_j q_{ij}n_j(t),\quad i=1,\ldots,l,
\end{equation}
or, in the matrix form
\begin{equation}\label{eq:1:2}
    \bs n(t+1)=\bs{QW}\bs n(t),\quad \bs n(t)=\bigl(n_1(t),\ldots,n_l(t)\bigl)^\top\in\R^l\,.
\end{equation}

Since the model \eqref{eq:1:1}, \eqref{eq:1:2} is linear it is possible to have three different outcomes: either the total population size will explode to infinity, or tend to zero, or, in some exceptional cases, will stay constant. From the evolutionary point of view we are mostly interested in the population composition and therefore it is natural to consider the system for the corresponding frequencies
$$
\bs p(t)=\frac{\bs n(t)}{\sum_{i=1}^l n_i(t)}\,,
$$
which takes the form
\begin{equation}\label{eq:1:3}
    \bs p(t+1)=\frac{\bs{QWp}(t)}{\overline w(t)}\,,
\end{equation}
where
$$
\overline{w}(t)=\sum_{i=1}^l w_i p_i(t)=\bs w\cdot \bs p(t)
$$
is \textit{the mean population fitness} that guarantees that $\bs p(t)$ is the probability distribution for any time moment $t$, that is it belongs to the simplex $S_l$ for any time moment:
$$
\bs p(t)\in S_l=\{\bs x\in \R^l\colon \sum_{i=1}^l x_i=1,\,x_i\geq 0\}.
$$
The dot denotes the usual dot product in $\R^l$.

Before moving forward we note that if we assume that no mutations occur the system becomes $p_i(t+1)=\frac{w_i}{\overline{w}(t)}p_i$, and hence
$$
\frac{p_i(t+1)}{p_{j}(t+1)}=\frac{w_i}{w_j}\frac{p_i(t)}{p_j(t)}\,,
$$
which immediately implies that, assuming without loss of generality that $w_1$ is the strict maximum of the fitness landscape, $p_1(t)\to 1$ and for all the rest $p_j(t)\to 0$. Moreover, the change in the mean population fitness is given by
$$
\Delta \overline w(t)=\overline w(t+1)-\overline w(t)=\frac{\sum_{i=1}^l \bigl(w_i-\overline w(t)\bigr)^2 p_i(t)}{\overline w(t)}=\frac{1}{\overline w(t)}\text{Var}_t(\bs w)\,,
$$
which is arguably the simplest form of Fisher's theorem of natural selection (see, e.g., \cite{burger2000mathematical} for the discussion and mathematical underpinnings of this ``theorem''). That is, the behavior in case of no mutations is very simple: all the types of individuals except for the most fit one are being washed out from the population; moreover, the mean population fitness is increasing at each step, and the magnitude of the increase is proportional to the variance of the fitness landscape at each time moment.

To be able to write an analogous system in continuous time one must separate the processes of selection (as described by the fitness coefficients) and mutation because, strictly speaking, only one elementary event can occur during a sufficiently small period of time. Hence, assuming that $\mu_{ij}$ is \textit{the mutation rate} of an individual of type $j$ into an individual of type $i$, and $m_i\in\R$ is the (\textit{Malthusian}) fitness of individuals of type $i$ (which is the difference of the birth and death rates and can be negative in this case), then the change of the population numbers is described by
$$
\bs{\dot n}(t)=\bs(\bs M+\bs{\mathcal M})\bs n(t),
$$
where we introduce the notations for the fitness landscape $\bs M=\diag(m_1,\ldots,m_l)$ and mutation matrix $\bs{\mathcal M}=[\mu_{ij}]$. Similarly to the above, it is more natural (see, e.g., \cite{karev2010} for the discussion) to consider the equations for the frequencies, which in this case take the form
\begin{equation}\label{eq:1:4}
    \bs{\dot p}(t)=\bigl(\bs M-\overline{m}(t)\bs I\bigr)\bs p(t)+\bs{\mathcal M}\bs p(t),
\end{equation}
where
\begin{equation}\label{eq:1:5}
    \overline{m}(t)=\sum_{i=1}^l m_i p_i(t)=\bs m\cdot \bs p(t)
\end{equation}
is the mean (Malthusian) fitness of the population and $\bs I$ is the identity matrix.

Model \eqref{eq:1:4}, \eqref{eq:1:5} was dubbed by Baake and co-authors as a \textit{paramuse} model, due to the parallel mutation selection scheme, see \cite{baake1999}. This model also got some treatment in an influential population genetic textbook by Crow and Kimura \cite{crow1970introduction}, and therefore is often called \textit{the Crow--Kimura model}.

Models \eqref{eq:1:3} and \eqref{eq:1:4} are intrinsically related since it can be shown \cite{hofbauer1985selection} that model \eqref{eq:1:4} is a limit for small generation time of model \eqref{eq:1:3}. The same limiting procedure helps relate the Wrightian and Malthusian fitnesses
$$
w_i=e^{m_i\Delta t}\approx 1+m_i\Delta t,\quad \Delta t\to 0,
$$
and the mutation probabilities and corresponding mutation rates
$$
q_{ij}=\delta_{ij}+\mu_{ij}\Delta t,\quad \Delta t\to 0,
$$
where $\delta_{ij}$ is the Kronecker delta.

For model \eqref{eq:1:4} also a version of the Fisher theorem of natural selection holds since, by elementary manipulations,
$$
\dot{\overline m}(t)=\text{Var}_t(\bs m)\geq 0,
$$
and similarly only one type of individuals survives in the long run (assuming as before that there is a strict maximum of the fitness landscape).

Both models \eqref{eq:1:3} and \eqref{eq:1:4} can be called the quasispecies models, but the fact is that Eigen in his 1971 paper considered yet another mathematical model, which takes the form (we use the notations introduced before, but now a care should be exercised since clearly in the continuous setting one needs to talk about rates and not probabilities)
\begin{equation}\label{eq:1:6}
    \bs{\dot p}(t)=\bs{QWp}(t)-\overline w(t)\bs p(t).
\end{equation}
We are not aware of any elementary mechanical derivation of \eqref{eq:1:6} but note that the equilibrium point of this model coincides with the fixed point of \eqref{eq:1:3}.

Mathematical models \eqref{eq:1:3}, \eqref{eq:1:4}, and \eqref{eq:1:6} share several common features. In particular,
\begin{enumerate}
\item Selection does not lead to a homogeneous population, as it happens in the systems with no mutations.

More precisely, under some mild technical conditions necessary to apply the Perron--Frobenius theory for nonnegative matrices, models \eqref{eq:1:3}, \eqref{eq:1:5}, and \eqref{eq:1:6} possess the only globally asymptotically stable equilibrium
$$
\lim_{t\to\infty} \bs p(t)=\bs p,
$$
which is the positive eigenvector of the eigenvalue problem
\begin{equation}\label{eq:1:7}
    \bs{QWp}=\lambda \bs p,
\end{equation}
for \eqref{eq:1:3} and \eqref{eq:1:6}, and of the eigenvalues problem
\begin{equation}\label{eq:1:8}
    (\bs{M}+\bs{\mathcal M})\bs p=\lambda \bs p,
\end{equation}
for \eqref{eq:1:4}. Moreover, this eigenvector corresponds to the dominant real eigenvalue of the matrices $\bs{QW}$ and $\bs M+\bs{\mathcal M}$ respectively, which is equal to the mean population fitness at the equilibrium $\bs p$:
$$
\lambda=\overline w=\bs w\cdot \bs p,\quad\text{or}\quad \lambda=\overline m=\bs m\cdot \bs p.
$$

This vector $\bs p$ was called by Eigen \textit{the quasispecies}, which is the target of selection in these models.

\item The mean population fitnesses $\overline w(t)$ or $\overline m(t)$ are not necessarily increasing functions of time; that is, the evolution in these quasispecies models does not imply the steady climbing of the fitness landscape. (This was first noted, to our knowledge, in \cite{schuster1988stationary}, where it was shown that it is possible to have (a) $\overline w(t)$ is non-decreasing along trajectories, (b) $\overline w(t)$ is non-increasing along trajectories, and $(c)$ $\overline w(t)$ may increase or decrease or even pass through extremum.) A detailed discussion and additional references to this quite frequent phenomenon can be found in \cite{bratus2017adaptive}.

\item For some specific fitness landscapes there exists a sharp transition of the equilibrium distribution $\bs p$ as a function of mutational landscape which separates the phase where the quasispecies vector is generally concentrates around the fittest type (the so-called selection phase) and the uniform distribution of the quasispecies (no selection, or random, phase). This transition was called \textit{the error threshold} and have clear connections to the phenomenon of \textit{phase transition} in statistical physics. We discuss this phenomenon in more details below.
\end{enumerate}

We emphasize that mathematically the asymptotic behavior of the quasispecies models boils down to the analysis of high-dimensional eigenvalue problems \eqref{eq:1:7} and \eqref{eq:1:8}. Due to the Perron--Frobenius theory we always know that the quasispecies distribution exists, it is a different, and much more complicated, question how to calculate $\bs p$ and the mean population fitness for some specific cases of fitness landscape and mutation matrix. In the rest of this review we survey various specific types of matrices $\bs{QW}$ and $\bs M+\bs{\mathcal M}$, for which at least partial information about the leading eigenvalues and the corresponding quasispecies eigenvectors can be obtained. Our presentation is necessarily biased since, especially in the second part of the review, we concentrate on our own results, which are discussed, together with detailed proofs, at length in \cite{bratus2013linear,semenov2014,semenov2016eigen,semenov2015,semenov2017generalized}.

\section{Sequence spaces and exact solutions}
To make progress in mathematical analysis of eigenvalue problems \eqref{eq:1:7} and \eqref{eq:1:8} one needs to specify the structure of the matrices involved. We deliberately did not specify what we understand by ``individual type'' in the previous section. In the language of population genetics the models considered above describe evolution in one-locus haploid population with $l$ alleles. Since Eigen was interested primarily in the problem of the origin of life, his interpretation of ``individual type'' was quite different. By analogy with RNA and DNA molecules he considered the population of \textit{sequences} of fixed length $N$, where each different sequence is composed from the letters of some finite alphabet. For example, one can take the four letter alphabet of nucleotides $\{A,T,G,C\}$, and therefore there will be $l=4^N$ different sequences in the population, or 20-letter alphabet of amino acids, and hence there will be $l=20^N$ different polypeptides. The simplest choice, however, is to consider initially two letter alphabet $\{0,1\}$ and hence deal with the population of binary sequences of fixed length $N$ having total $l=2^N$ different sequence types.

\begin{figure}[!ht]
\centering
\includegraphics[width=0.49\textwidth]{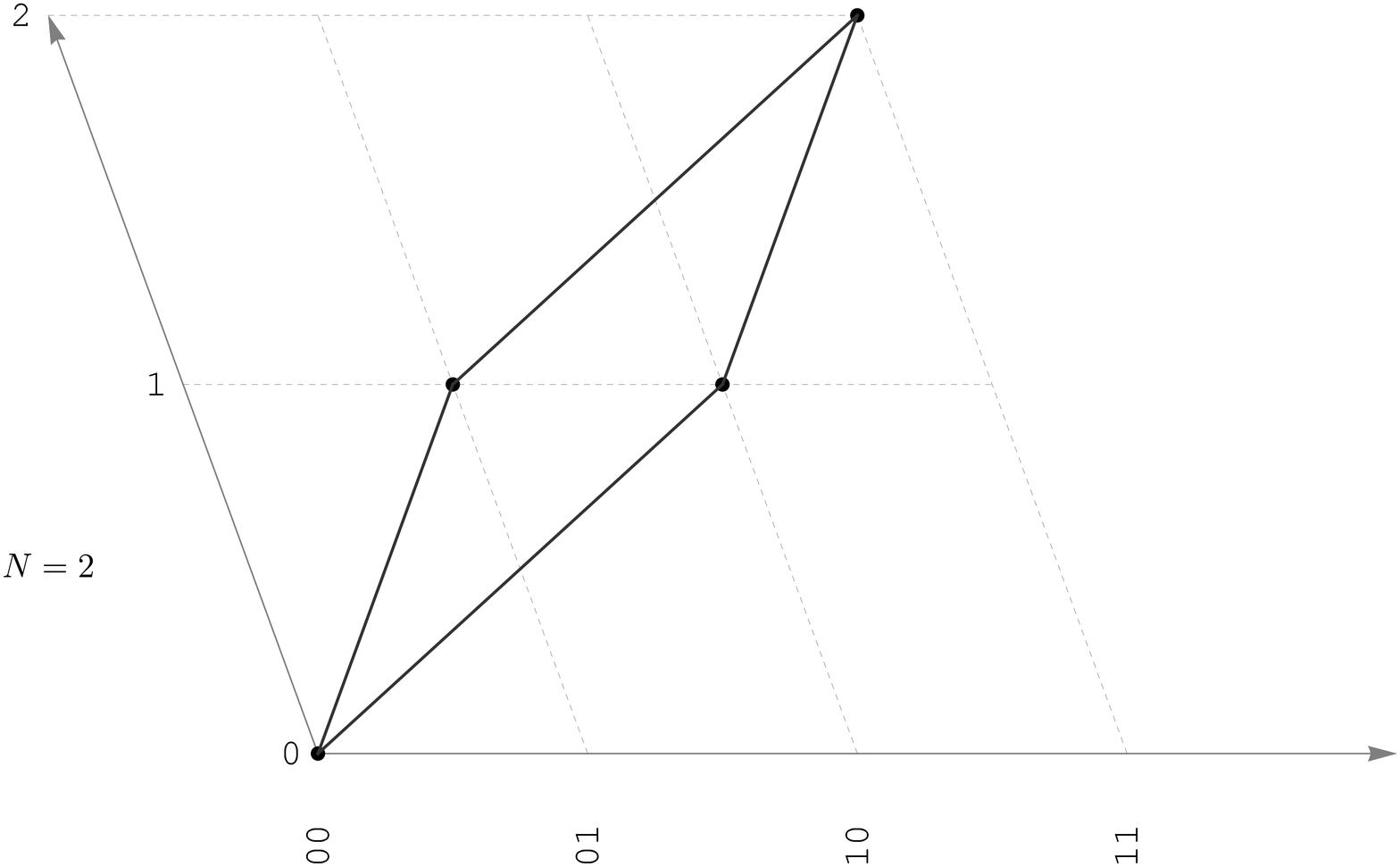}
\includegraphics[width=0.49\textwidth]{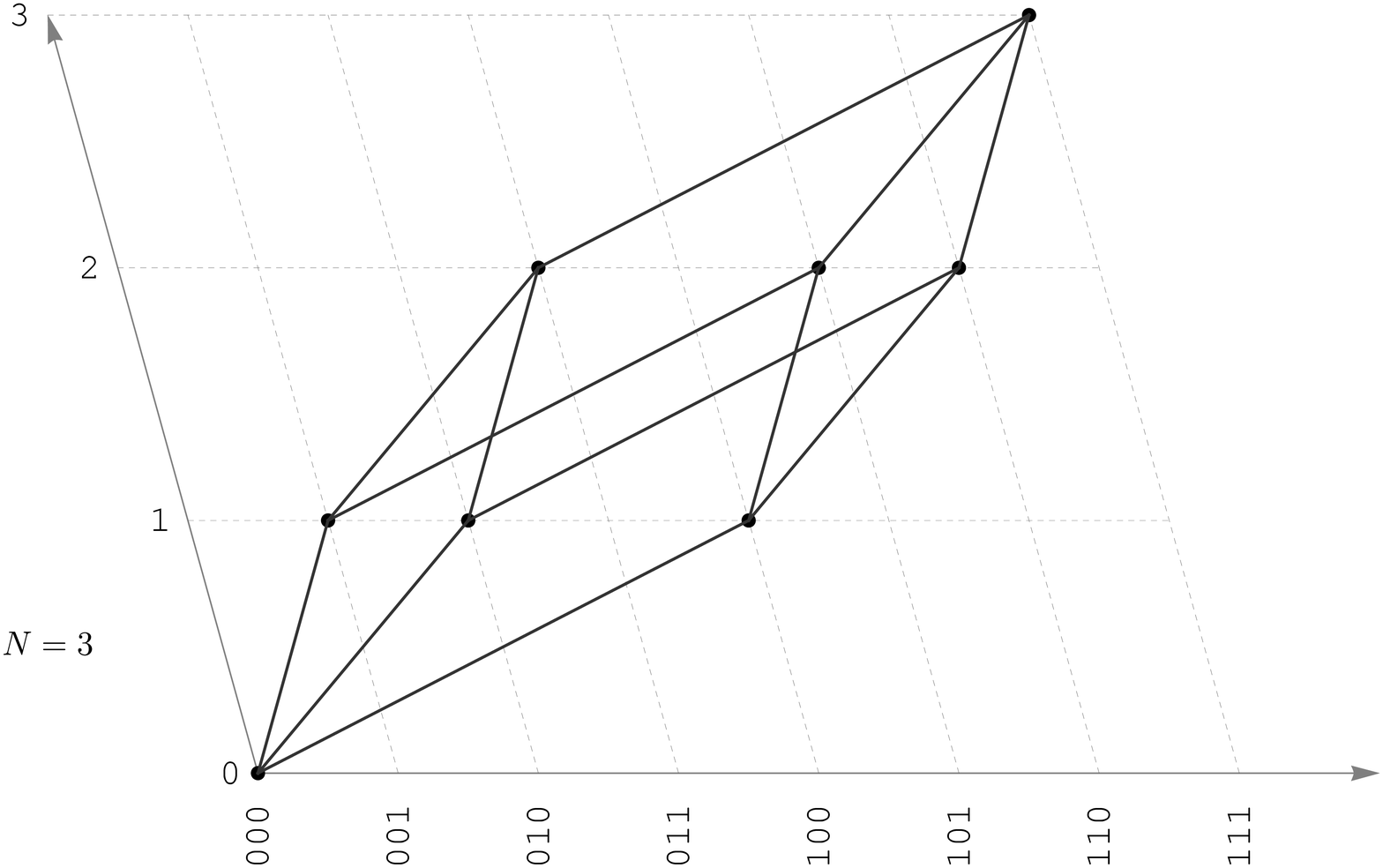}\\
\hfill

\includegraphics[width=0.49\textwidth]{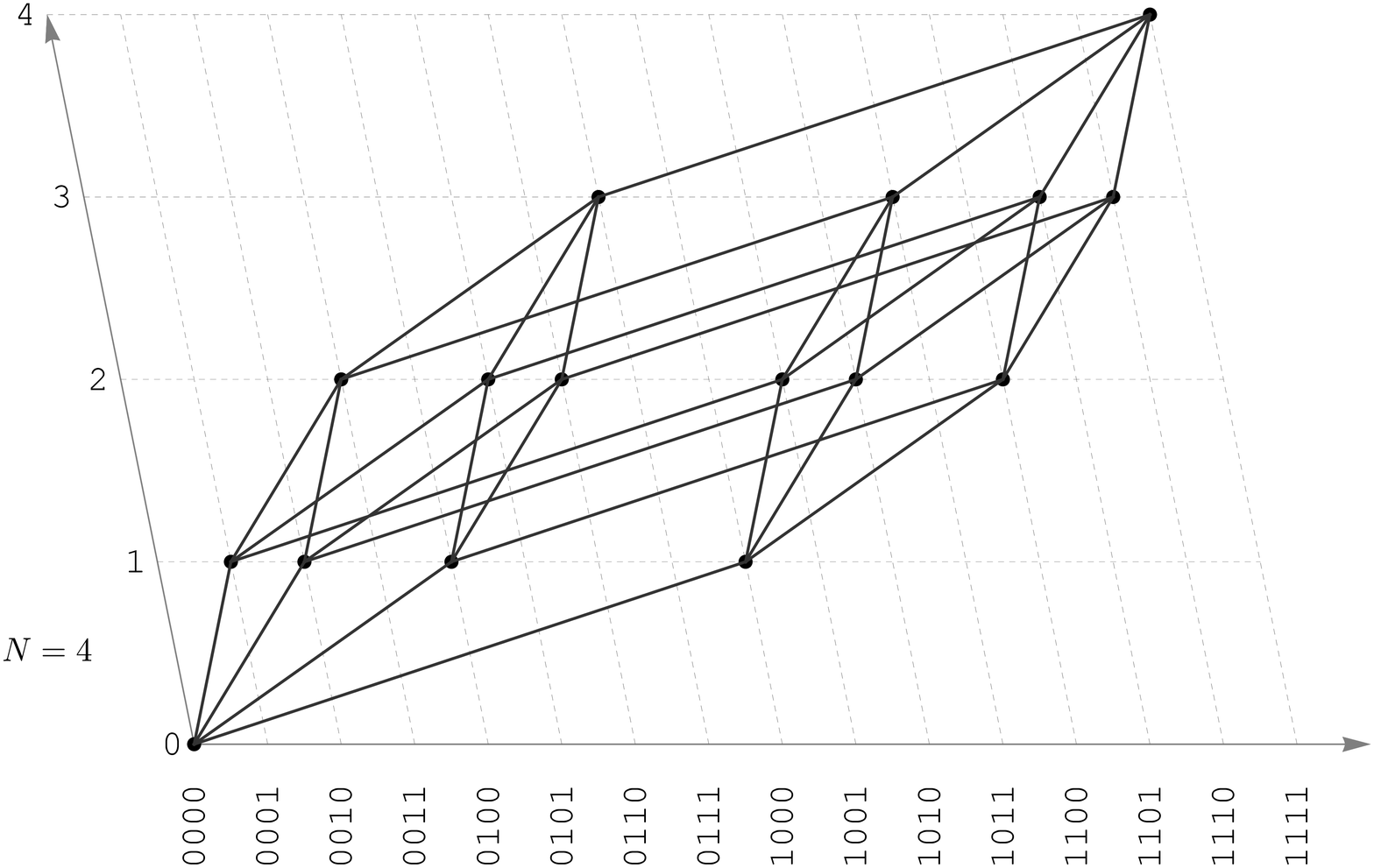}
\includegraphics[width=0.49\textwidth]{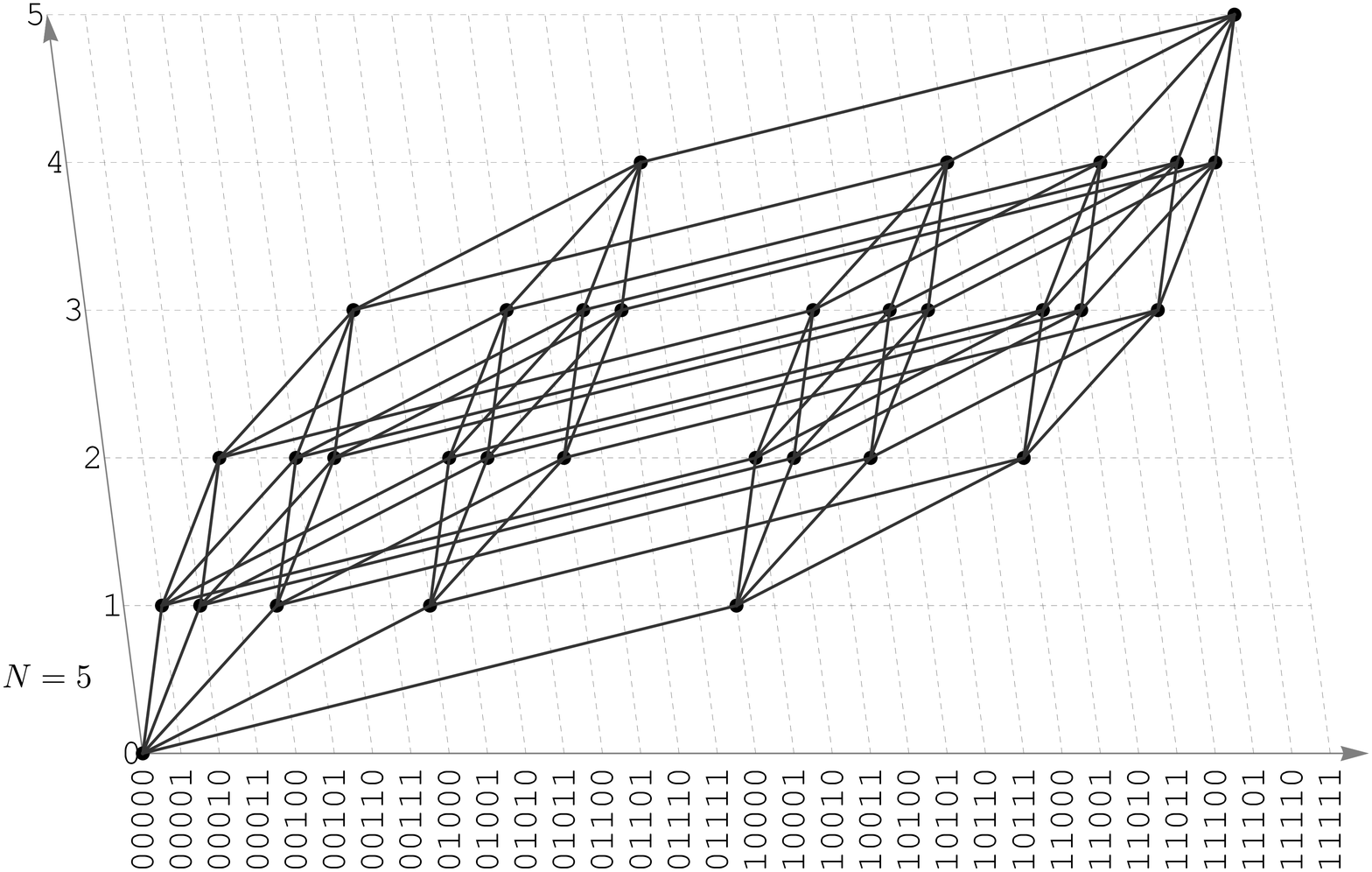}
\caption{The binary sequence spaces for sequences of length $N=2,3,4,5$. The vertices of the hypercube correspond to various sequences, which are listed along the $x$-axis, and the Hamming distance between any two sequences is equal to the least number of edges connecting the corresponding vertices.}\label{fig:1}
\end{figure}

Such underlying space of binary sequences  possesses a nice geometric interpretation: Different types of binary sequences of length $N$ are in one-to-one correspondence with the vertices of an $N$-dimensional hypercube (see Fig. \ref{fig:1}).

Moreover, now we can, under some additional assumptions, describe in much more detail the corresponding mutation matrices. Namely, assume that mutations occur independently at each site of the sequences with the same probability $q$ and introduce the Hamming distance $H_{ij}$ between sequences of types $i$ and $j$. For the following it is convenient to use the lexicographical order of different sequences such that sequence of type $i$, where $i=0,\ldots,2^N-1$, has exactly the binary representation of integer $i$, supplemented, if necessary, by additional zeros.
Now, the elements of the mutation matrix $\bs Q$ are given by
\begin{equation}\label{eq:2:1}
    q_{ij}=(1-q)^{N-H_{ij}}q^{H_{ij}},\quad i,j=0,\ldots, 2^N-1,
\end{equation}
that is, matrix $\bs Q$ is defined in terms of only one scalar parameter $q$.

In continuous time (model \eqref{eq:1:4}) the mutations during short time interval are only possible to the neighboring sequences, and therefore the mutation rates are given by
\begin{equation}\label{eq:2:2}
    \mu_{ij}=\begin{cases}\mu,& H_{ij}=1,\\
    -N\mu,& H_{ij}=0,\\
    0,& H_{ij}>1.
    \end{cases}
\end{equation}

Using the introduces sequence spaces, the eigenvalue problems \eqref{eq:1:7} and \eqref{eq:1:8} now depend only on the fitness landscapes $\bs W$ and $\bs M$ and mutation parameters $q$ and $\mu$ respectively, and hence it is customary to emphasize this dependence by writing $\bs p(q)$ and $\overline w(q)$ or $\bs p(\mu)$ and $\overline m(\mu)$ for the equilibrium quasispecies distributions and mean population fitnesses. We emphasize that in the rest of this review we talk exclusively about the stationary distributions and do not discuss any time-dependent aspects of the quasispecies evolution.

\section{Permutation invariant fitness landscapes and error threshold}
We posed in the previous section the main mathematical question regarding the quasispecies theory, namely, how, given matrices $\bs W$ or $\bs M$, give an exact and/or approximate expressions for leading eigenvalue and, if possible, the corresponding positive eigenvector $\bs p$ of the eigenvalue problems \eqref{eq:1:7} and \eqref{eq:1:8} with the mutation matrices \eqref{eq:2:1} and \eqref{eq:2:2} respectively as the functions of the mutation probability $q$ and the mutation rate $\mu$. One of the main reasons why this question turned out so complicated in general is the possibility of the phenomenon, which was called \textit{the error threshold}. First we illustrate this phenomenon numerically, following the simplifications that were originally introduced in \cite{swetina1982self}.

Note that even if the alphabet we use to compose our sequences has only two letters, the total number of different types of sequences is $2^N$, which becomes unrealistically large for numerical computations even for modest values of $N$. Swetina and Schuster \cite{swetina1982self}, to overcome this difficulty, suggested to use the so-called \textit{single peak fitness landscape}, in which one of the sequences, called the master sequence, is assigned the higher value of the fitness constant, and everyone else is ``equally inferior,'' that is
$$
\bs w=(w+s,w,w,\ldots,w),\quad \text{or}\quad \bs{m}=(m+s,m,\ldots,m),
$$
where constant $s>0$ is the selective advantage of the master sequence. In both cases the master sequence was chosen to be composed of all zeros.

The single peak fitness landscape is an example of what we call \textit{permutation invariant fitness landscapes}, for which the fitness of the given sequence is determined not by the sequence itself but by the total number of ones in this sequence (or, more mathematically rigorous, by the Hamming norm of the sequence, which is, by definition, is the distance from this sequence to the zero sequence, $H_i:=H_{0i}$). In this way the total population is divided now into \textit{classes} of sequences, which are characterized by the Hamming distance from the zero sequence, such that $j$-th class contains exactly $N\choose j$ types of sequences. Now, if we are able to modify our mutation matrix accordingly, then the dimension of the problem is reduced from $2^N\times 2^N$ to $(N+1)\times (N+1)$ since there are exactly $N+1$ different classes of binary sequences of length $N$.

The easiest way to do it is for the continuous time model \eqref{eq:1:8}, \eqref{eq:2:2}. Indeed, let $\nu_{ij}$ be the rate of mutations from class $j$ to class $i$. Assuming again that $\mu$ is the rate of mutations per one site per time unit, we have
$$
\nu_{ij}=\begin{cases}(N-j)\mu,& i=j+1,\\
j\mu,& i=j-1,\\
-N\mu,& i=j,\\
0,& \text{otherwise},
\end{cases}
$$
since only one elementary event is possible per small time interval, and mutation in any of the 0 sites yields a sequence in class $j+1$, whereas mutation in any of the 1 sites yields a sequence in class $j-1$.

Therefore, for the permutation invariant fitness landscapes and Crow--Kimura quasispecies model \eqref{eq:1:4} the eigenvalue problem \eqref{eq:1:8}, \eqref{eq:2:2} takes the form
\begin{equation}\label{eq:3:1}
    (\bs M+\bs{\mathcal N})\bs p=\lambda \bs p,
\end{equation}
where, as before, $\bs M=\diag(m_0,\ldots,m_N)$ (we abuse the notations here, by using the same letter to denote now the fitness of the $j$-th class, a more correct, and uglier, notation would be $m_{H_j}$), and $\bs{\mathcal N}=[\nu_{ij}]_{(N+1)\times (N+1)}$, which has an especially simple tri-diagonal form
\begin{equation}\label{eq:3:2}
    \bs{\mathcal N}=\mu \bs S=\mu\begin{bmatrix}
                         -N & 1 & 0 & 0 & \ldots & \ldots & 0 \\
                         N & -N & 2 & 0 & \ldots & \ldots & 0 \\
                         0 & N-1 & -N & 3 & \ldots & \ldots & 0 \\
                         \ldots & \ldots & \ldots & \ldots & \ldots & \ldots & \ldots \\
                         0 & 0 & \ldots & \ldots & 2 & -N & N \\
                         0 & 0 & \ldots & \ldots & 0 & 1 & -N \\
                       \end{bmatrix}\,.
\end{equation}

For the eigenvalue model \eqref{eq:1:7}, \eqref{eq:2:1} the bookkeeping of the mutation probabilities is slightly more involved and leads \cite{nowak1989error} to the eigenvalue problem
\begin{equation}\label{eq:3:3}
    \bs{WRp}=\lambda\bs p,
\end{equation}
with $\bs W=\diag(w_0,\ldots,w_N)$, the fitness landscape of sequence classes, and matrix $\bs R=[r_{ij}]$, were $r_{ij}$ is the probability that a sequence from class $j$ mutates into sequence of class $i$, explicitly \cite{nowak1989error}
\begin{equation}\label{eq:3:4}
    r_{ij}=\sum_{a=j+i-N}^{\min\{i,j\}}{j \choose a}{N-i \choose j-a}q^N\left(\frac{1-q}{q}\right)^{i+j-2a}\,,\quad i,j=0,\ldots, N.
\end{equation}

\begin{figure}[!b]
\centering
\includegraphics[width=0.48\textwidth]{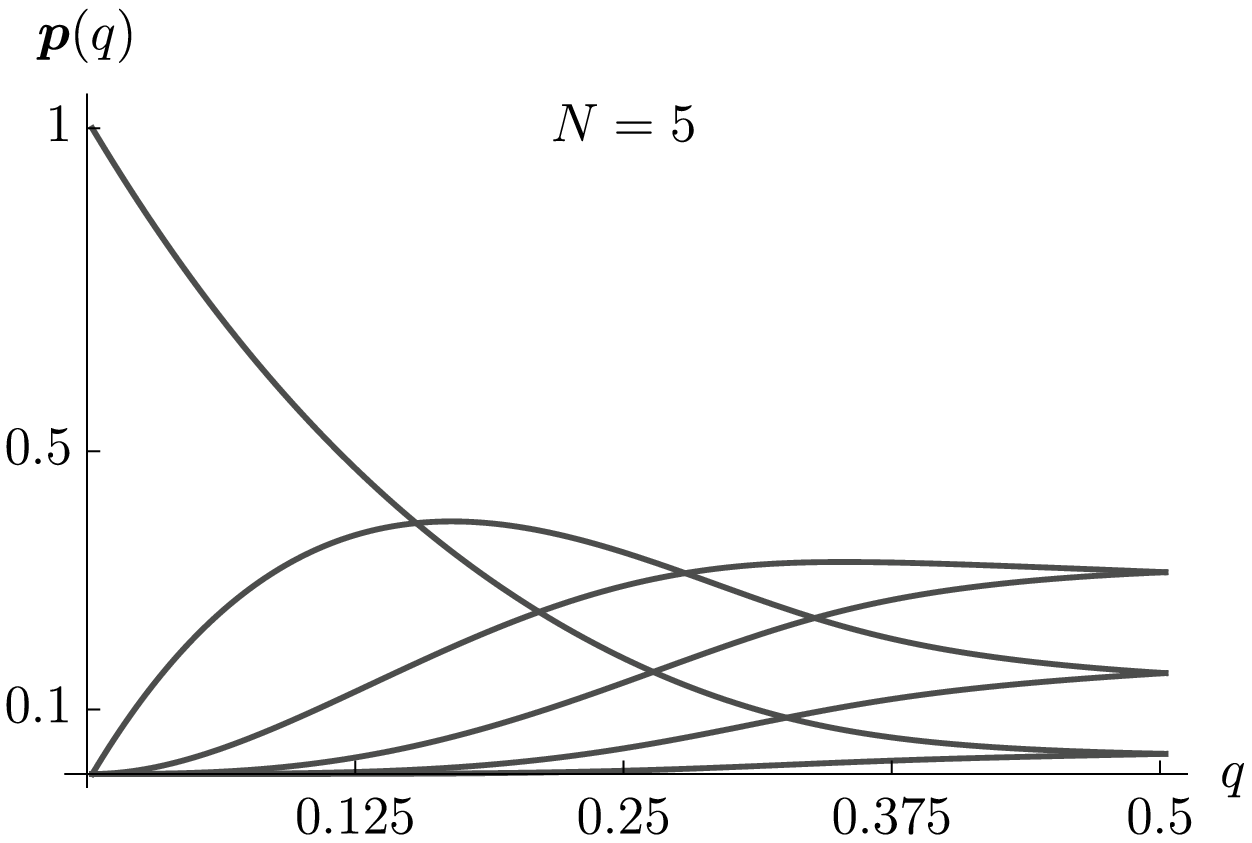}\hfill
\includegraphics[width=0.48\textwidth]{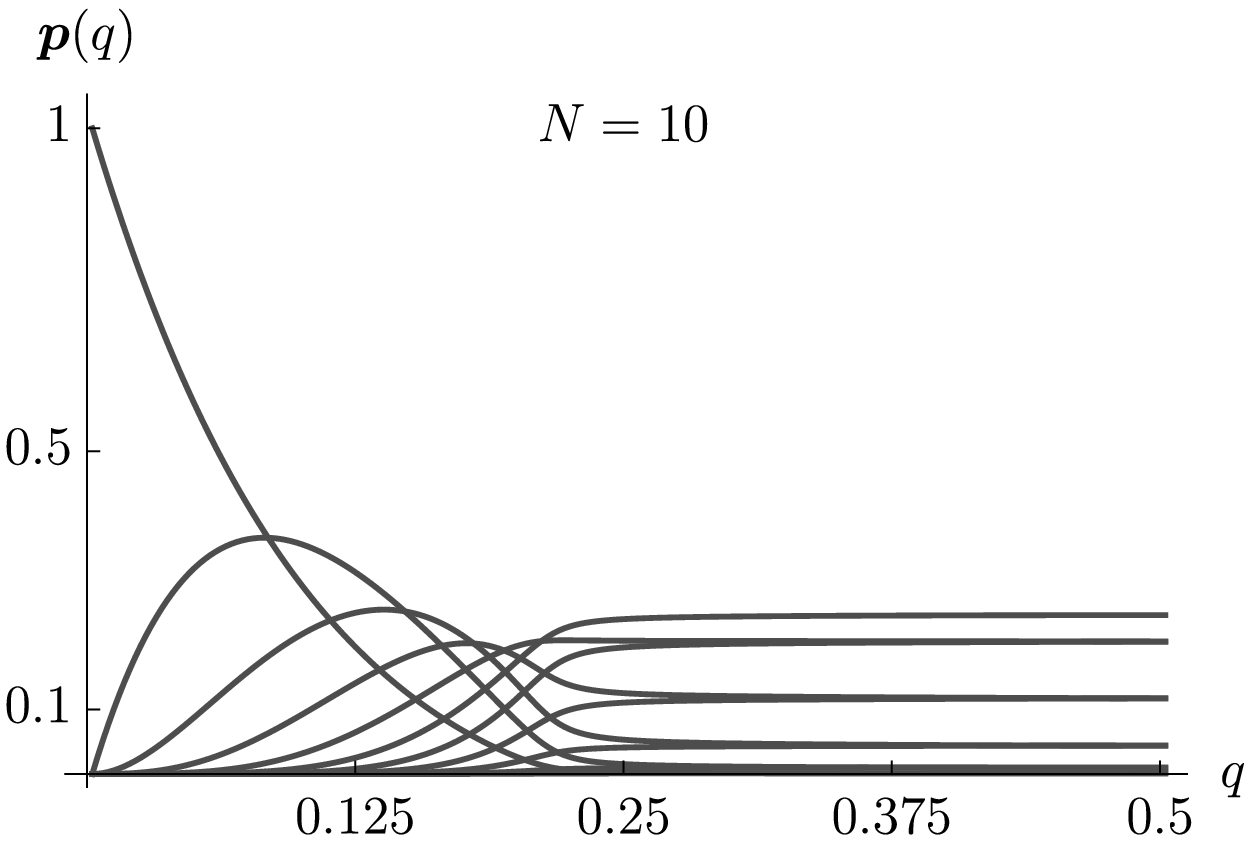}\\[4mm]
\includegraphics[width=0.48\textwidth]{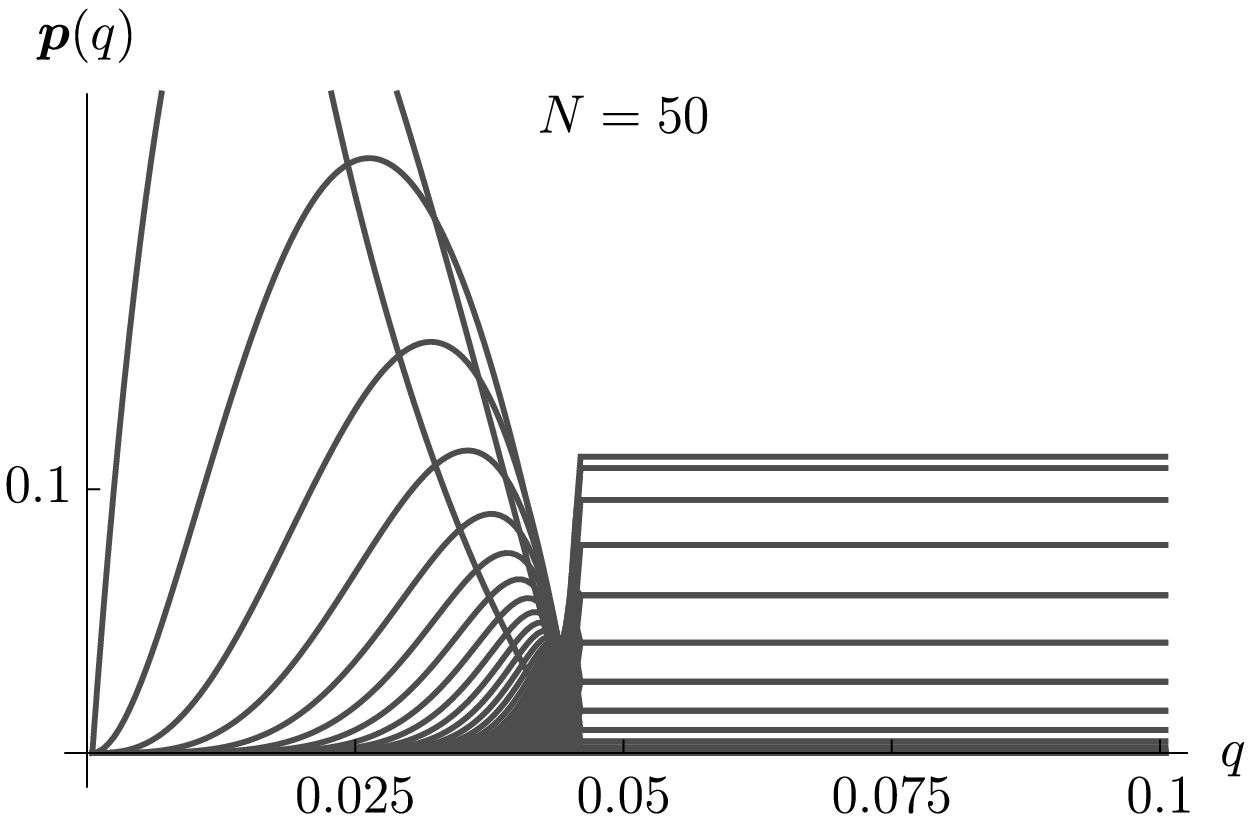}\hfill
\includegraphics[width=0.48\textwidth]{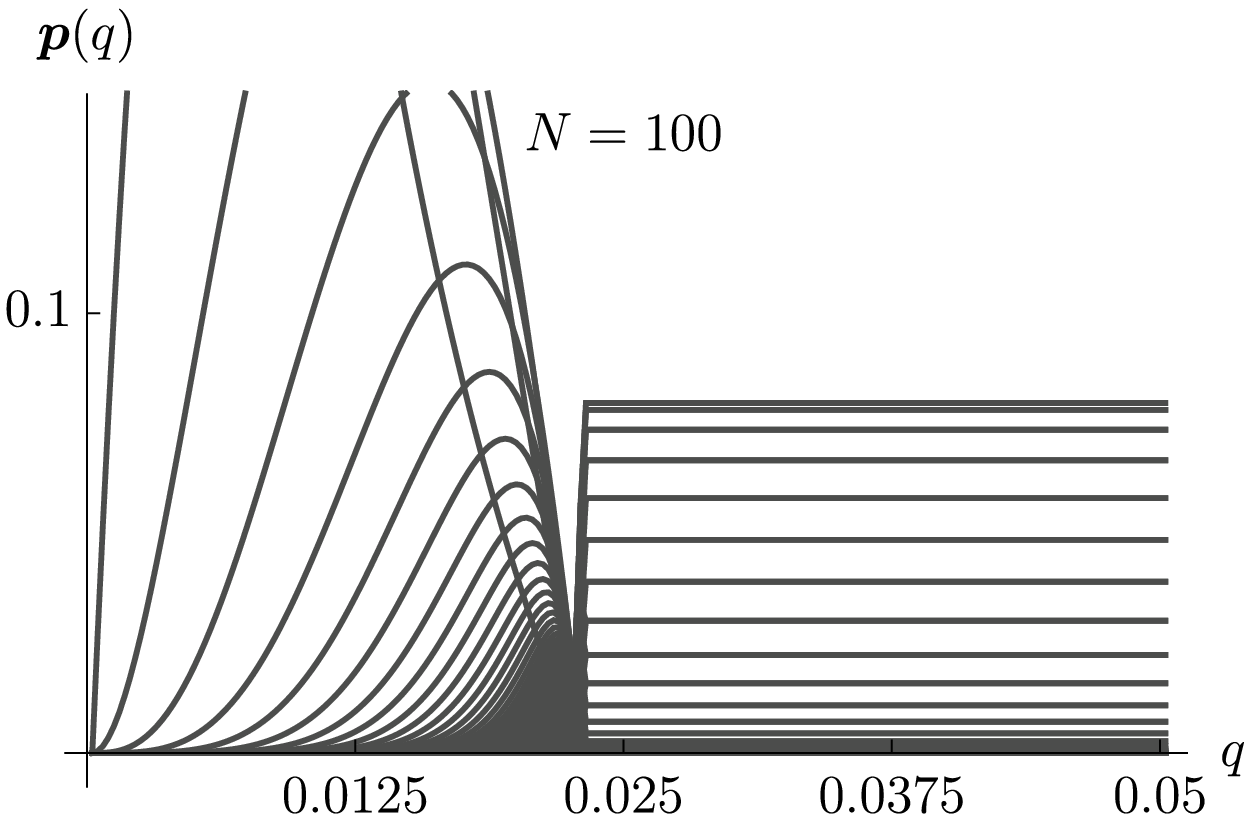}
\caption{The quasispecies vector of the eigenvalue problem \eqref{eq:3:3}, \eqref{eq:3:4} depending on the mutation probability $q$ per site for different sequence lengthes $N$. The fitness landscape is $\bs W=\diag(10,1,\ldots,1)$. Note the different scales for different graphs. }\label{fig:2}
\end{figure}
Here is a numerical illustration how the equilibrium quasispecies distribution $\bs p(q)$ changes with respect to $q$ for different sequence lengthes and the single peak fitness landscape $\bs W=\diag(10,1,\ldots,1)$, see Fig. \ref{fig:2}. It is quite clear that if the sequence length increases there appears a threshold value of the mutation probability, after which the quasispecies distribution of sequence classes does not change and remains binomial, which means that the distribution of the sequence types is (almost) uniform after this critical mutation rate. This phenomenon is called \textit{the error threshold}. Right at this point we note that this particular phenomenon is landscape dependent, and at least for some fitness landscapes does not manifest itself \cite{wiehe1997model}.

There is a heuristic way to derive the value of the error threshold for the given $w,s,N$. Basically one assumes that no backward mutations to the fittest class are possible, and then the first differential equation for the master sequence takes the form
$$
\dot p_0(t)=\bigl((w+s) r_{00}-\overline{w}(t)\bigr )p_0(t),\quad \overline w(t)=w+s p_0(t).
$$
We have $r_{00}=(1-q)^N$, and hence the condition
$$
(w+s)(1-q)^N=w
$$
implies ``extinction'' of the master sequence (we put the word ``extinction'' into the quotes since, strictly speaking, the quasispecies models are written for the frequencies and hence it makes no sense to discuss the phenomenon of extinction within the framework of these models). The last equality can be manipulated as
$$
q^\ast=1-\sqrt[N]{\frac{w}{w+s}}\,,
$$
which for the example above for $N=100$ gives the threshold value
$$
q^\ast\approx 0.0227,
$$
which is very close to the sharp transition observed in Fig. \ref{fig:2}. The same relation can be manipulated as follows
\begin{equation}\label{eq:3:5}
N^\ast=\frac{\log \sigma}{q}\,,\quad \sigma=\frac{w+s}{w}\,,
\end{equation}
which is often interpreted as the critical condition on the attainable length of a polynucleotide sequence given the mutation probability. Again, jumping ahead, we note that there exists a rigorous proof of the formula \eqref{eq:3:5}. We note, however, that this formula applies \textit{only} to the case of the single peak landscape, and therefore it is dangerous to make conclusions that sequence length is inversely proportional to the mutation probability. In \cite{semenov2014} we conjectured that for the general quasispecies model \eqref{eq:1:7}, \eqref{eq:2:1} the error threshold mutation rate $q^\ast$, if it exists, can be determined by the formula
$$
1-q^\ast=\sqrt[N]{\frac{\overline w(0.5)}{\overline w(0)}}=\frac 12\sqrt[N]{\frac{\sum_{i=0}^{2^N-1}w_i}{\max_j\{w_j\}}}\,.
$$
Among other things, this expression shows that the inverse relationship of the sequence length and critical mutation probability does not pertain to any possible fitness landscape, for more details see \cite{semenov2014}.

Very similar picture is observed for the Crow--Kimura quasispecies model. In particular, the formula \eqref{eq:3:5} turns into
$$
N^\ast=\frac{s}{\mu}\,,
$$
for the single peak landscape $\bs M=(m+s,m,\ldots,m),\,s>0$. We reiterate that it is generally not applicable to other possible fitness landscapes.

\section{The only exact solution and isometry group of the hypercube}
Here we return back to the general eigenvalue problems \eqref{eq:1:7}, \eqref{eq:2:1} and \eqref{eq:1:8}, \eqref{eq:2:2}, i.e., we do not assume here that the fitness landscape is permutation invariant. The discussed above  phenomenon of the error threshold indirectly implies that it is quite naive to expect that even for such simple fitness landscapes, such as single peak landscape, we expect to find simple explicit formulas for the mean population fitness and quasispecies distribution. There is, however, one special case, for which exact solution can be written for a specific general fitness landscape. This observation is due to Rumschitzki \cite{rumschitzki1987spectral}, who
noticed that in general for the mutation matrix $\bs Q$ can be found simple decomposition
$$
\bs Q_k=\bs Q_1\otimes \bs Q_{k-1},\quad k=2,\ldots, N,
$$
where
$$
\bs Q_1=\begin{bmatrix}
          1-q & q \\
          q & 1-q \\
        \end{bmatrix},
$$
and $\otimes$ is the Kronecker or tensor product of matrices (e.g., \cite{laub2005matrix}). This decomposition immediately implies that all the eigenvalues and eigenvector of matrix $\bs Q$ can be found in terms of easily analyzed matrix $\bs Q_1$. To include also the fitness landscape, let us define the following matrices
$$
\bs W_k=\begin{bmatrix}
          1 & 0 \\
          0 & s_k \\
        \end{bmatrix},\quad k=1,\ldots,N.
$$
If we define recursive procedure
$$
(\bs Q\bs W)_k=(\bs Q_1\bs W_k)\otimes (\bs {QW})_{k-1},\quad k=2,\ldots, N,
$$
where $(\bs{QW})_1=\bs Q_1\bs W_1$, then at the $N$-th step we obtain that
$$
(\bs{QW})_N=\bs Q\bs W,
$$
where $\bs Q$ is given by \eqref{eq:2:1} and matrix $\bs W$ is diagonal, with elements
$$
w_{ii}=\prod_{k\colon a_k=1}s_{k},\quad i=1,\ldots, 2^N-1,\quad w_{00}=0,
$$
assuming that index $i$ has the binary representation
$$
i=[a_1,a_2,\ldots,a_N],\quad a_k\in\{0,1\}.
$$
Since the matrix $\bs{QW}$ is given again as $N$-fold Kronecker product then all the eigenvalues and eigenvectors can be calculated using the eigenvalues and eigenvectors of $2\times 2$ matrices $\bs Q_1\bs W_k$.

Biologically this exactly solvable case corresponds to the multiplicative fitness landscape, when the fitness of a given sequence is represented by the multiplicative (independent) contribution of all the sites with 1's. It is customary in this case to speak of \textit{no epistasis} fitness landscape, that is, the cumulative effect of all the sites is given by independent contributions of each site.

Similar formulas can be written for the Crow--Kimura model \eqref{eq:1:8} with \eqref{eq:2:2}, in this case the fitness landscape becomes additive, that is, again, fitness of a given sequence is given by independent contributions of each site. All these formulas can be directly generalized to more then two letters in the sequence alphabet and to non-uniform mutation probabilities (rates) along the sequences, providing a false sense of generality. What is more important however, and what was clearly observed in \cite{dress1988evolution}, the potential solvability of the quasispecies model is directly related to the isometry group of the underlying finite metric space, which can be identified with $N$-dimensional hypercube, presented in Fig. \ref{fig:1}. It is interesting to note that the connections of the Eigen eigenvalue problem and the isometry group of the hypercube were not explored further until very recently.

\section{The Ising model, the maximum principle, and the Hamilton--Jacobi equation approach}
In the same year when the paper by Rumischitzki \cite{rumschitzki1987spectral} was submitted for publication, it was noted in \cite{leuthausser1986exact} that the eigenvalue problem \eqref{eq:1:7}, \eqref{eq:2:1} was already studies for one specific fitness landscape in the disguise of the transfer matrix of two-dimensional Ising model of statistical physics \cite{onsager1944crystal}. On one hand this observation emphasized the nontriviality and complexity of analytical analysis of the quasispecies model, on the other hand it opened the gaits for a stream of papers that used statistical physics methods to analyze different reincarnations of the quasispecies model (see, e.g., \cite{baake1997ising,galluccio1997exact,leuthausser1987statistical}, and especially \cite{baake2001mutation} and references therein). The methods borrowed from statistical physics usually imply some infinite sequence limit under an appropriate scaling of the model parameters, and hence the results are mostly asymptotical, contrary to the exact results can be obtained with the formulas from the previous section. In line with the tradition from statistical physics these asymptotical results are very often called ``exact'' in the literature, which should be kept in mind while studying such approaches. The methods of statistical physics indeed allowed significant progress in the analysis of the quasispecies model, as it is discussed at length in a very detailed and accessible paper \cite{baake2001mutation}. Most importantly, they caused a number of researches to formulate and prove, by rigorous mathematical methods, what is now called \textit{the maximum principle} \cite{Baake2005,Baake2007,Hermisson2002}.

The maximum principle applies to the models \eqref{eq:3:1}, \eqref{eq:3:2} and \eqref{eq:3:3}, \eqref{eq:3:4}, that is, to the permutation invariant fitness landscapes, and allows to obtain, under some technical conditions, an approximation for the mean population fitness. For model \eqref{eq:3:1}, \eqref{eq:3:2} it takes the following form (we note that we do not try to formulate the most general form of the maximum principle): Assume that the permutation invariant fitness landscape can be represented as
$$
m_i=N f(x_i),\quad x_i=\frac{i}{N}\in[0,1],
$$
and define function $g(x)=\mu-2\mu\sqrt{x(1-x)}$. Then it can be proved that the mean population fitness $\overline{m}(\mu)$ satisfies
\begin{equation}\label{eq:4:1}
    \overline{m}(\mu)\approx N\sup_{x\in[0,1]}\bigl(f(x)-g(x)\bigr)\,.
\end{equation}
For example, assume that we deal with single peak landscape of the form
$$
\bs M=\diag(N,0,0,\ldots,0),
$$
hence $f(x)=1$ if $x=0$ and $f(x)=0$ if $x>0$. Simple calculations imply that in this case
$$
\overline m(\mu)=\begin{cases}N(1-\mu),&\mu<1,\\
0,&\mu\geq 1,
\end{cases}
$$
which also provides a proof of the threshold mutation rate for the single peak landscape discussed earlier. This approximation provides a remarkable agreement with numerical computations (see, e.g., \cite{bratus2013linear}). An analogous result holds for the classical model \eqref{eq:3:3}, \eqref{eq:3:4}. It was obtained initially in \cite{saakian2006ese}, and rigorous proof can be found in \cite{Baake2007}.

It is important to emphasize at this point that although the maximum principle is an extremely powerful device to analyze various fitness landscapes, it only applies to permutation invariant landscapes, and also relies on several technical conditions, the most important of which is the continuity of function $f$: more precisely, in \cite{Hermisson2002} it was assumed that $f$ must have only finite number of discontinuities and be either left or right continuous at every point; in \cite{Baake2007} the continuity of $f$ was required. We are still far from understanding the realistic general fitness landscapes, but it is almost universally accepted that in many cases the evolution proceeds with huge leaps, and at least some fitness landscapes are essentially discontinuous. There are simple examples (see below) that show that a formal application of the maximum principle in the case of discontinuous fitness landscapes can lead to erroneous conclusions (\cite{bratus2013linear,wolff2009robustness}).

Another mathematical approach to the quasispecies models with permutation invariant fitness landscapes is based on the limiting procedure that transform the original system of ordinary differential equations into one first order partial differential equation of Hamilton--Jacobi type. There is a recent review of results obtained with this approach \cite{saakian2015mathematical}, therefore here we just mention the main idea. Again, for simplicity, we only present the results for model \eqref{eq:3:1}, \eqref{eq:3:2}.

In terms of total population numbers the Crow--Kimura model with tri-diagonal mutation matrix can be written as
$$
\dot n_i(t)=m_i n_i(t)-N\mu n_i(t)+(N-i+1)\mu n_{i+1}(t)+(i+1)\mu n_{i-1}(t),\quad i=0,\ldots, N.
$$
Now we introduce the ansatz
$$
n_i(t)=e^{Nu(t,x)},\quad x=1-\frac{2i}{N}\,,\quad m_i=Nf(x),
$$
and use formal Taylor series. After dropping the terms of higher order with respect to $1/N$ we obtain the equation
$$
\frac{\partial u}{\partial t}=f(x)-\mu+\mu\frac{1-x}{2}e^{2\frac{\partial u}{\partial x}}+\mu\frac{1+x}{2}e^{-2\frac{\partial u}{\partial x}}\,,\quad -1\leq x\leq 1,
$$
for some given initial condition $u(0,x)$.

We are not aware of any rigorous proof that would justify the discussed above limit; we note however that assuming that this equation indeed holds in the limit of infinite sequence length it is quite straightforward to obtain the maximum principle \eqref{eq:4:1} \cite{wolff2009robustness}, which implies that at least some continuity conditions of $f$ should be required.

\section{Linear algebra of the quasispecies model}
Having discussed the maximum principle and the Hamilton--Jacobi equation, we remark that in both approaches first some infinite sequence limit is taken and after it the problem at hands is analyzed. In the papers \cite{bratus2013linear,semenov2014} we undertook a somewhat different approach, in which we first rigourously manipulate the corresponding eigenvalue problems and only after we take the limit $N\to\infty$. It turns out that in this way it is possible to obtain new results, which cannot be derived by the maximum principle or the Hamilton--Jacobi equation.

As we discussed above it is quite straightforward to calculate the eigenvalues and eigenvectors of mutation matrices \eqref{eq:2:1}, \eqref{eq:2:2}, \eqref{eq:3:2}, and \eqref{eq:3:4}. Moreover, as these calculations show, the corresponding eigenvectors always form a basis for $\R^{l}$ or $\R^{N+1}$. Therefore, the basic idea that we used, was to rewrite the full eigenvalue problem in the basis of the eigenvectors of the mutation matrices.

In particular, in \cite{bratus2013linear} we analyzed the problem \eqref{eq:3:1} and found that it is possible to derive a parametric solution to this problem in the form
\begin{equation}\label{eq:5:1}
    p_i(s)=F_{ij}(s),\quad F(s)=m_j F_{jj}(s),\quad \mu=\frac{s}{2}F(s),\quad \overline m(s)=F(s),
\end{equation}
where $s$ is some parameter,
$$
F_{ij}(s)=2^{-N}\sum_{k=0}^N\frac{c_{ik}c_{kj}}{1+ks}\,,
$$
and matrix $\bs C=[c_{ij}]$ is composed of the eigenvectors of matrix $\bs S$ defined in \eqref{eq:3:2}. It may look that this specific parametric form  of the solution to the eigenvalue problem \eqref{eq:3:1} does not simplify the situation. It turns out, however, that, given an explicit form of the fitness landscape $\bs M$, these formulas allow us to make further analysis, and in particular, consider, under an appropriate scaling, the limiting procedure $N\to\infty$. Here are two examples, the proofs can be found in \cite{bratus2013linear}.
\begin{proposition}Consider the eigenvalue problem \eqref{eq:3:1}, \eqref{eq:3:2}.

If the fitness landscapes is the single peak landscape $\bs M=\diag(N,0,\ldots,0)$ then, for $\mu<1$,
\begin{equation}\label{eq:5:2}
\lim_{N\to \infty} \frac{\overline{m}(\mu)}{N}=1-\mu,\quad \lim_{N\to\infty}p_i(\mu)=(1-\mu)\mu^i,\quad i=0,\ldots,N.
\end{equation}

If the fitness landscape is $\bs M=\diag(0,\ldots,0,N,0,\ldots,0)$, where $N=2A$ and the only non-zero rate exactly at the position $A$, then
\begin{equation}\label{eq:5:3}
\lim_{N\to\infty}\frac{\overline{m}(\mu)}{N}=\overline{r}_\infty=\sqrt{\mu^2+1}-\mu,\quad p_{A\pm k}\approx \left(\frac{1-\overline{r}_\infty}{1+\overline{r}_\infty}\right)^k.
\end{equation}
\end{proposition}

Several remarks are in order. First, we do not claim that the result in \eqref{eq:5:2} is new. To the best of our knowledge, the expression for $\overline{m}/N$ was derived originally in \cite{galluccio1997exact} and, as we showed above, elementary follows from the maximum principle. In \cite{saakian2004solvable} the geometric distribution of the quasispecies vector was derived using some heuristic methods. We provided a rigorous proof illustrating our parametric solution, and also gave an estimate of speed of convergence to this distribution.
\begin{figure}[!t]
\centering
\includegraphics[width=0.46\textwidth]{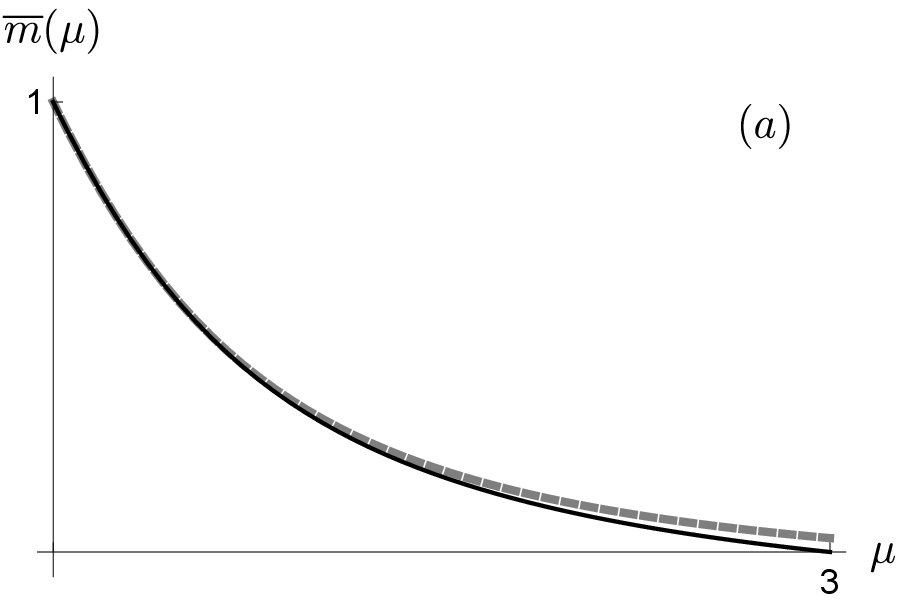}\hfill
\includegraphics[width=0.46\textwidth]{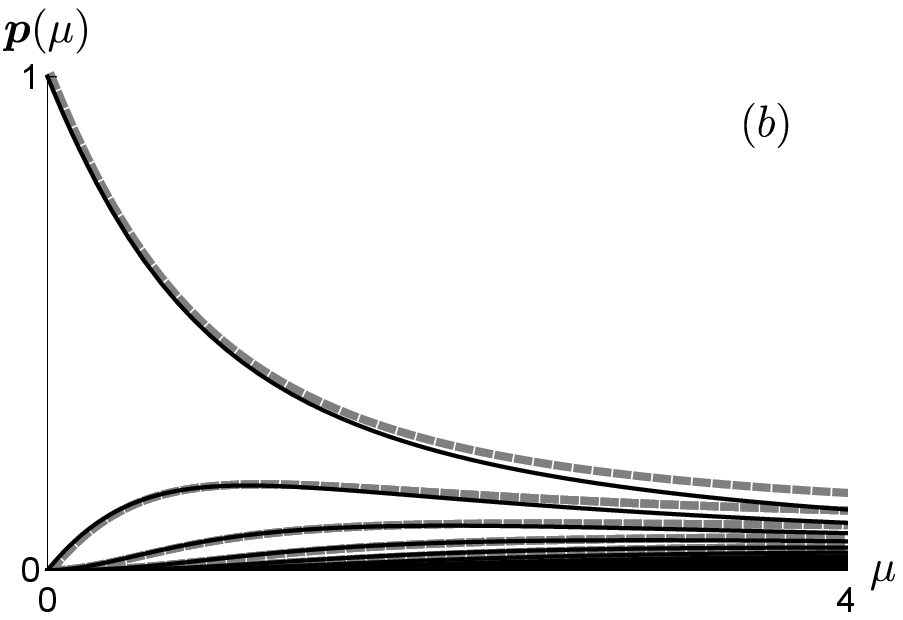}
\caption{Comparison of the numerical solutions (grey, dashed) of the eigenvalue problem \eqref{eq:3:1} and approximations in \eqref{eq:5:3} (black) for $N=100$. $(a)$ Mean population fitness, $(b)$ the quasispecies vector of frequencies of sequence classes.}\label{fig:3}
\end{figure}

Second, the example \eqref{eq:5:3} is \textit{not} a single peak landscape because here the whole class of sequences has the same fitness $N$, recall that we consider the case of permutation invariant fitness landscape. In Fig. \ref{fig:3} it can be seen that approximation is quite good even for moderate values of $N$. Also, as our analysis predicted, in this case there is no error threshold. Finally, this example cannot be analyzed by both the maximum principle and Hamilton--Jacobi approach, because the limit fitness function is essentially discontinuous at $x=1/2$ (neither left or right continuous), which shows that the approach first manipulate the eigenvalue problem and only after it take the limit $N\to\infty$ is in some sense more general.

We obtained some other interesting results based on parametric solution \eqref{eq:5:1}, but probably the most important consequence of its analysis was a heuristic approach that allowed actually compute the quasispecies distribution for a wide variety of fitness landscapes. We discuss this approach in the next section.

We undertook a similar approach in \cite{semenov2014} to analyze model \eqref{eq:1:7}, \eqref{eq:2:1}. As a result we obtained a number of proofs for the results whose validity was generally acknowledged, however, this acknowledgement was based mostly on numerical computations. We note that $q$ as defined in \cite{semenov2014} is equal to $1-q$ as defined in Section 2. In particular we prove \cite{semenov2014}
\begin{theorem}\label{th:1}Consider the eigenvalue problem \eqref{eq:1:7}, \eqref{eq:2:1}. If the fitness landscape $\bs W$ is such that the leading eigenvalues $\overline{w}(0),\overline w(1)$ as functions of the mutation probability $q$ have multiplicities 1, then there exist an absolute minimum $\hat{w}$ of function $\overline w(q)$ for $0.5\leq q<1$. The point of this minimum is determined by the condition $\overline w'(q)=0$. For $q\leq 0.5$ function $\overline w(\hat q)$ is nondecreasing and convex.
\end{theorem}
\begin{figure}[!b]
\centering
\includegraphics[width=0.45\textwidth]{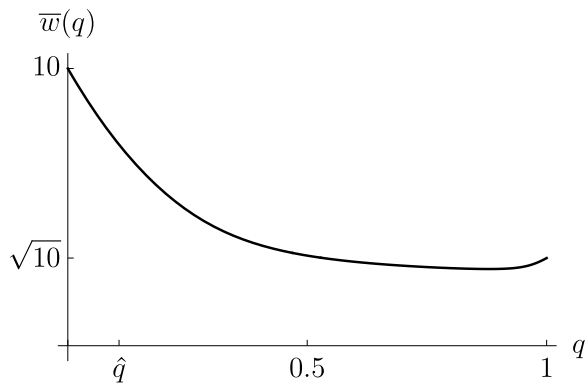}
\caption{A numerical illustration of Theorem \ref{th:1}: A typical picture of the qualitative behavior of $\overline{w}(q)$. Here $N=3$, the fitness landscape is $\bs W=\diag(10,3,3,2,3,2,2,1)$, $\overline w(0)=10,\overline{w}(1)=\sqrt{10}$.}\label{fig:4}
\end{figure}

There are a number of other exact results in \cite{semenov2014}, but, similar to the previously discussed work, probably the most important result of this work was the general idea how to advance an analysis of \textit{non} permutation invariant fitness landscapes. We discuss these ideas in Sections \ref{sec:7} and \ref{sec:8}.
\section{Formulas for the quasispecies distribution}\label{sec:6}While the maximum principle provides a powerful tool to analyze the behavior of the mean population fitness, there are very few results of explicit computations of the quasispecies distribution, an example \eqref{eq:5:2} notwithstanding. In \cite{semenov2015} two of us suggested a general heuristic procedure how to derive a limiting quasispecies distribution for the model \eqref{eq:3:1}.

In \cite{semenov2015} we noticed that the matrix $\bs S$ in \eqref{eq:3:2} is exactly the matrix in the standard polynomial basis of the linear differential operator
$$
\mathcal S\colon P(s)\mapsto (1-s^2)P'(s)-N(1-s)P(s),
$$
and hence the eigenvalue problem \eqref{eq:3:2} can be written in the form
$$
\bs m\circ P(s)+\mu(1-s^2)P'(s)-\mu N(1-s)P(s)=\overline{m}P(s),
$$
where $\bs m\circ P(s)=\sum_{i=0}^N m_ip_is^i$.
By dividing by $N$ and taking the formal limit $N\to \infty$ we end up with
\begin{equation}\label{eq:6:1}
    \bs r\circ P(s)-\mu (1-s)P(s)=\overline rP(s),\quad \bs r=\frac{\bs m}{N}\,,\quad \overline r=\frac{\overline m}{N}\,.
\end{equation}
Now we conjecture (but did not prove rigorously) that if for some given $\bs r$ one can solve equation \eqref{eq:6:1} with the conditions $\bs r\circ P(1)=\overline r,P(1)=1$ then $P(s)$ is the probability generating function of the equilibrium quasispecies vector.

For example, if we take $\bs r=(1,0,\ldots,0)$, that is, the single peak landscape, then $\bs r\circ P(s)=P(0)$ and hence \eqref{eq:6:1} takes the form
$$
-\mu(1-s)P(s)+P(0)=\overline r P(s).
$$
Plugging $s=0$ into the last expression and assuming $P(0)\neq 0$ we find immediately
$$
\overline r=1-\mu,
$$
that is the same expression as in \eqref{eq:5:2}. Using the condition $P(1)=1$ yields immediately
$$
P(s)=\frac{1-\mu}{1-\mu s}\,,
$$
which is the probability generating function of the geometric distribution with the parameter $\mu$ (see again \eqref{eq:5:2}).

\begin{figure}
\centering
\includegraphics[width=0.45\textwidth]{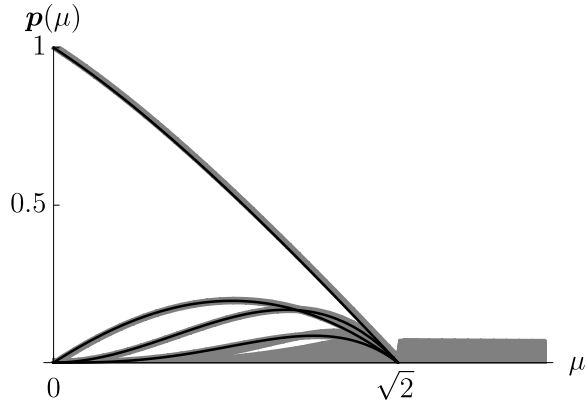}
\caption{A numerical illustration of Lemma \ref{l:1}: Numerical solution (grey) with $N=200$ is compared with the analytical predictions for the fitness landscape $\bs m=N(2,0,1,0,1,\ldots)$. Note the error threshold and its theoretical prediction $\sqrt{2}$.}\label{fig:5}
\end{figure}

Assuming that our approach valid we can prove the following general fact:
\begin{lemma}\label{l:1}Assume that $\bs r=(r_0,r_1,\ldots)$ such that $r_0>r_i,\, i=1,2,\ldots$. Then
$$
\overline r=r_0-\mu,\quad p_i=\frac{\mu^ip_0}{\prod_{j=1}^i(r_0-r_j)}\,,\quad i>0,\quad p_0=\frac{1}{1+\sum_{k=1}^\infty \frac{\mu^k}{\prod_{j=1}^k (r_0-r_j)}}\,.
$$
These formulas provide a solution for the quasispecies distribution in the limit case $N\to\infty$ if and only if
$$
\mu<\mu^\ast=\liminf \sqrt[n]{\prod_{j=1}^n (r_0-r_{j})}.
$$
The critical value $\mu^\ast$ gives the error threshold of the mutation rate of the permutation invariant Crow--Kimura model.
\end{lemma}
For example for the fitness landscape $\bs r=(2,0,1,0,1,0,\ldots)$ we immediately conclude that the error threshold we have
$$
\mu^\ast=\sqrt{2},
$$
and
$$
p_0=\frac{2-\mu^2}{2+\mu}\,,\quad p_1=p_0\frac{\mu}{2}\,,\quad\ldots,
$$
see the numerical illustration in Fig. \ref{fig:5}.

We finish this section noting that in a series of recent papers \cite{cerf2016quasispecies,cerf2016quasispeciesnew,dalmau2017asymptotic} Cerf and Dalmau analyzed a more complicated case of model \eqref{eq:3:3} and \eqref{eq:3:4}. As a result they obtained, in the form of infinite series, explicit solution for the quasispecies distribution for the permutation invariant Eigen quasispecies model \eqref{eq:3:3} and \eqref{eq:3:4}.

\section{Two-valued fitness landscapes and isometry group of the hypercube}\label{sec:7}
Let us reiterate that now we have very powerful tools to analyze permutation invariant fitness landscapes with the help of, e.g., the maximum principle or the explicit formulas discussed in the previous section. Much less is available to tackle non permutation invariant fitness landscapes, which are, of course, much more biologically realistic. In \cite{semenov2016eigen} two of us considered a special case of the full eigenvalue problem \eqref{eq:1:7}, \eqref{eq:2:1} with the fitness landscape that we dubbed \textit{two-valued fitness landscape}.

Let $A$ be a non-empty subset of indices $A\subseteq \{0,1,\ldots,l-1\}$. We fix two constants $w\geq 0$ and $s>0$ and consider the fitness landscape of the form
$$
w_k=\begin{cases}
w+s,&k\in A,\\
w,&k\notin A.
\end{cases}
$$

One of the main theoretical results can be then formulated as follows (for a proof, which is based on the analysis in \cite{semenov2014}, see \cite{semenov2016eigen}).
\begin{lemma}Let $X$ be the finite metric space on indices $\{0,1,\ldots,2^N-1\}$ with the Hamming distance between the binary representations of indices (i.e., geometrically the $N$ dimensional hypercube, see Fig. \ref{fig:1}), let $G$ be a group that acts on $X$ by isometries (i.e., $G\leq \text{Iso}(X)$), and let $A$ in the definition of the two-valued fitness landscape be a $G$-orbit. Then the mean population fitness $\overline{w}(q)$ is a root of algebraic equation (with coefficients depending on $q$) of degree at most $N+1$.
\end{lemma}

We note for the readers that are not closely familiar with the language of a group action, in \cite{semenov2016eigen} there is a section with completely elementary discussion of the terminology and results.

Also we remark that we give an explicit form for the algebraic equation for $\overline{w}(q)$. Here are just two examples.

First, let us consider again the classical single peak landscape $\bs W=\diag (w+s,w,\ldots,w)$. The group $G$ here is simply the trivial group. Then the equation that determines $\overline{w}$ is given by
$$
\frac{1}{2^N}\sum_{d=0}^N {N\choose d}\frac{(2q-1)^d}{\overline w-w(2q-1)^2}=\frac{1}{s}\,.
$$
A very similar expression for a slightly different model was obtained originally in \cite{galluccio1997exact}.

Of course, single peak landscape is an example of a permutation invariant fitness landscape. Here is an example that is not permutation invariant.

Let
$$
G=Q_8=\left\{\pm 1,\pm i,\pm j,\pm k \mid i^2=j^2=k^2=-1,ij=k,jk=i,ki=j\right\}
$$
be the classical quaternion group of order 8. Consider the embedding $Q_8\to S_8$ where we choose $i\to(0212)(4657),j\to (0415)(2736)$. As a $G$-orbit we can, for instance, take
$$
A=\{7,11,13,14,112,176,208,224\}\subset X,\quad N\geq 8.
$$
It can be seen immediately that this fitness landscape is not permutation invariant. We can show that, e.g., for $N=8$, the leading eigenvalue of \eqref{eq:1:7}, \eqref{eq:2:1} is determined by
$$
\sum_{d=0}^8\frac{R_d(2q-1)}{\overline w-w(2q-1)^d}=\frac{64}{s}\,,
$$
where
$$
R_0=R_8=2,R_1=R_7=1,R_2=R_6=14,R_3=R_5=15,R_4=0.
$$
Equally easy to write an algebraic equation for an arbitrary $N$.

In the nutshell, the geometry of the underlying hypercube turned out to be crucial in determining the cases, when significant simplification of the original $2^N\times 2^N$ eigenvalue problem can be made. This fact was originally noticed in \cite{dress1988evolution}, but did not get subsequent development until our recent paper \cite{semenov2016eigen}.
\section{Abstract quasispecies model}\label{sec:8}
Having studied the two-valued fitness landscapes discussed in the previous section, we asked a very natural mathematical question: Why to focus all the attention on the hypercube? From mathematical point of view nothing precludes us from considering the following \textit{generalized quasispecies model} \cite{semenov2017generalized}.

Let $(X,d)$ be a finite metric space with integer valued metric $d$. Consider a group $\Gamma\leq \text{Iso}(X)$ of isometries of $X$ that acts transitively on $X$. Since it acts transitively, we can pick any point $x_0\in X$ and consider the function $d_{x_0}\colon X\longrightarrow \N$ such that $d_{x_0}(x)=d(x_0,x)$. By definition, the diameter of $X$ is $N=\text{diam}\, X=\max\{d_{x_0}(x)\mid x\in X\}$. Also consider a fitness function $\bs w\colon X\longrightarrow \R_{\geq 0}$ which can be represented as a vector with non-negative components. Together with the introduced notations consider fitness matrix $\bs W=\diag(\bs w)$, symmetric mutation matrix $\bs Q=[q^{d(x,y)}(1-q)^{N-d(x,y)}]$ for $q\in[0,1]$, and the distance polynomial
$$
p_X(q)=\sum_{x\in X}q^{d(x,x_0)}(1-q)^{N-d(x,x_0)},\quad x_0\in X.
$$
Now we call the problem to find the leading eigenvalue $\overline w(q)$ and/or the corresponding positive eigenvector $\bs p$ of the eigenvalue problem
$$
\bs{QWp}=p_X(q)\overline w\bs p
$$
the generalized algebraic quasispecies problem. It turns into classical Eigen's problem \eqref{eq:1:7}, if $X=\{0,1\}^N$ is the $N$-dimensional binary cube with the usual Hamming metric, in this case $p_X(q)=1$.

We gave an extensive treatment of the generalized quasispecies problem in \cite{semenov2016eigen,semenov2017generalized}, here we would like to state just one specific result. Namely, consider a generalization of two-valued fitness landscape in the form that the fitness function $\bs w$ is constant on each $G$-orbit, where $G\leq \Gamma$ and has at least two values. Consider the decomposition
$$
X_0=A_0\sqcup\bigsqcup_{i=1}^t A_i,
$$
such that $A_0$ is the union of $G$-orbits on which $\bs w(A_0)=w\geq 0$, and $\bs w(A_i)=w+s_i,s_i>0$.  Then the following theorem can be proved (see \cite{semenov2017generalized}).

\begin{theorem}The dominant eigenvalue of all the examples of the generalized quasispecies eigenvalue problem considered in \cite{semenov2017generalized} can be found as a root of an algebraic equation of degree at most $t\cdot (N+1)=t\cdot ({\rm diam}(X)+1)$. Moreover, this equation can be written down in the explicit form.
\end{theorem}

\begin{figure}
\centering
\includegraphics[width=0.8\textwidth]{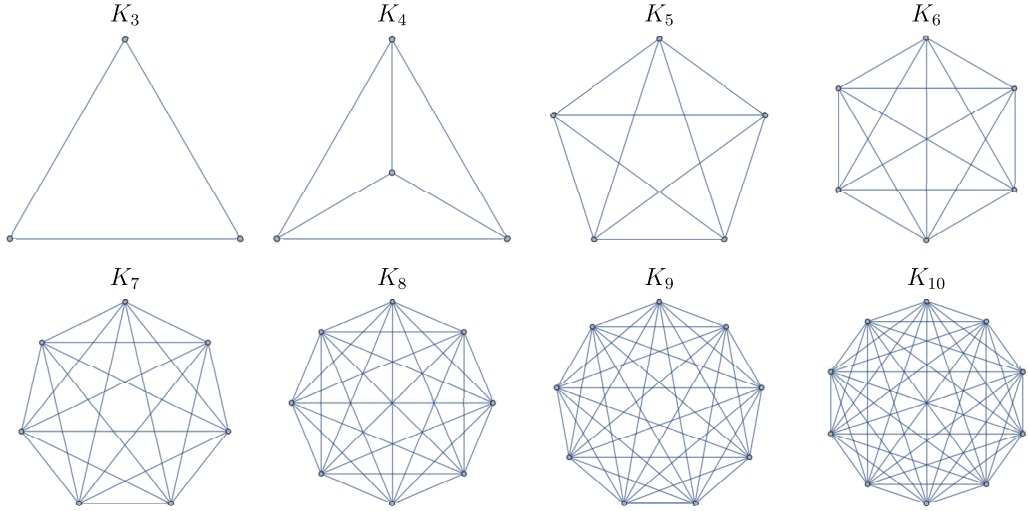}
\caption{Complete graphs on 3 to 10 vertices, representing the mutational landscape with simplex geometry.}\label{fig:6}
\end{figure}

Arguably the simplest possible generalized quasispecies model is generated by the geometry of simplex, which can be represented as a complete graph (see Fig. \ref{fig:6}). In this case we deal with a finite metric space $X$ of diameter $N=1$, that is we consider the case in which individuals of a population can mutate into any other individual with the same probability $q$. We note that this abstraction can be actually considered as a mathematical description of of the switching of antigenic variants for some bacteria.
Let $A\subseteq X$ such that $\bs w(A)=w+s$. Then, according to the theorem above, the leading eigenvalue of the generalized quasispecies problem can be found as a root of a quadratic equation, which (see \cite{semenov2016eigen} for a derivation) takes the form
$$
\frac{|A|}{(n+1)(\overline u-u)}+\left(1-\frac{|A|}{n+1}\right)\frac{2q-1}{(q+n(1-q))\overline u-(2q-1)u}=1,\quad u=\frac{w}{s}\,,\quad \overline u=\frac{\overline w}{s}\,.
$$
Here $|A|$ is the cardinality of set $A$, and $n=|X|$. That is, this case turns out to be significantly simpler then the classical quasispecies problem. Moreover, it can be proved that for this simplicial landscape the error threshold exists (see the discussion and rigorous derivations in \cite{semenov2016eigen}).

In a similar manner other possible geometries can be analyzed, see, for instance, the analysis of regular $m$-gon and hyperoctahedron mutational landscapes in \cite{semenov2017generalized}.

\section{Concluding remarks}
As we discussed above, since Manfred Eigen's original paper \cite{eigen1971sma} a lot of mathematical peculiarities of the quasispecies model were rigorously analyzed. At the same time we would like to conclude our presentation with a number of still open mathematical questions.
\begin{itemize}
\item In Section \ref{sec:6} we presented a heuristic algorithm to compute the quasispecies distribution for the permutation invariant Crow--Kimura model. A proof of validity of this approach is still missing.

\item There exist a few sufficient conditions on the fitness landscape for the error threshold to exist (e.g., \cite{wolff2009robustness}). We are not aware of any necessary and sufficient conditions of this sort.

\item While the abstract results on the generalized quasispecies model discussed in Section \ref{sec:8} are of significant interest, our paper \cite{semenov2017generalized} considers only the examples in which $X=A_0\sqcup A_1$. It is important to consider examples with more complicated partition of $X$ (for instance, important mesa-landscapes \cite{wolff2009robustness} have exactly this more complicated form).

\item In terms of the generalized quasispecies model it would be interesting to study the following question: What are the properties of $X$ and the distance function $d$ that guarantee that at least for some fitness landscapes the error threshold exists. This question also has some direct connections with various forms of the Ising model.
\end{itemize}

A list of open mathematical questions about the now classical quasispecies model can be easily extended, and it is our hope that at least for some of these problems methods and approaches discussed in this review can be of some help.


\end{document}